\documentclass[conference]{IEEEtran}
\usepackage{amsmath,amssymb}

\usepackage{comment}
\usepackage{amssymb}
\usepackage{bm}
\usepackage[dvipdfmx]{graphicx}

\graphicspath{{figure/}}

\newtheorem{defi}{Definition}

\usepackage{setspace}

\usepackage{algorithm}
\usepackage{algorithmic}
\usepackage{color}

\begin{document}
\title{Index ARQ Protocol for Reliable Contents Distribution over Broadcast Channels} 

\author{
  \IEEEauthorblockN{Takahiro Oshima and Tadashi Wadayama}
  \IEEEauthorblockA{Department Computer Science and Engineering, \\Nagoya Institute of Technology,
    Nagoya,  Japan\\
    Email: oshima@it.cs.nitech.ac.jp, wadayama@nitech.ac.jp} 
}

\maketitle
\begin{abstract}

In the present paper, we propose a broadcast ARQ protocol based on the concept of index coding. In the proposed scenario, a server wishes to transmit a finite sequence of packets to multiple receivers via a broadcast channel with packet erasures until all of the receivers successfully receive all of the packets. In the retransmission phase, the server produces a coded packet as a retransmitted packet based on the side-information sent from the receivers via feedback channels. A notable feature of the proposed protocol is that the decoding process at the receiver side has low decoding complexity because only a small number of addition operations are needed in order to recover an intended packet. This feature may be preferable for reducing the power consumption of receivers. The throughput performance of the proposed protocol is close to that of the ideal FEC throughput performance when the erasure probability is less than $0.1$. This implies that the proposed protocol provides almost optimal throughput performance in such a regime.
\end{abstract}

\section{Introduction}

Recent strong demand for handing data sets of extremely large size has produced a number of situations in which a server must send a huge file to multiple clients over an unreliable channel. A simple example is the distributed backup of a mission-critical file system. In order to avoid a devastating incident due to a natural disaster, distributed backups at distant locations are of critical importance. It is common to use multicast protocols to reduce the bottleneck traffic at the server for such applications. Since IP multicast is based on the User Data Protocol (UDP), which is not 
so reliable over an IP network, an IP multicast protocol, such as a reliable multicast protocol, 
is proposed in order to ensure reliable content distribution.
Another possible scenario is content distribution, such as HD video streams in cellular wireless systems. Suppose that a base station (i.e., server) wishes to share a large file or bitstream 
with multiple mobile terminals. In such a case, careful design 
of a protocol is necessary in order to achieve sufficient throughput while maintaining a certain degree of reliability.

These situations can be abstracted as a problem of sharing identical content with multiple receivers over an unreliable broadcast channel. A broadcast channel is a channel over which receivers can listen to what a server has sent over common media: wireless signals on a specific band or a multicast network. In order to achieve high reliability (i.e., low error probability) of data and high throughput, several coding techniques and protocols have been proposed for broadcast channels \cite{IBRP}\cite{SRB}\cite{MHA}\cite{RCRMDD}\cite{RMDD}\cite{PLRMT}.

A prominent example is Automatic Repeat reQuest (ARQ) protocols, which are often used for reliable content distribution over a broadcast channel. In an ARQ protocol, a receiver sends a request for retransmission to the sever if a receiver receives a broken packet or detects a packet loss. The server resends the corresponding packets when it receives a request from a receiver. Simple protocols for single-to-single communication, such as a Go-Back-$N$ protocol and a selective-repeat protocol, can also be used for broadcast channels. However, direct application of such protocols to a broadcast channel often causes significant degradation of the throughput performance because such protocols do not consider packet losses in distinct receivers.

On the other hand, a specialized ARQ protocol using Forward Error Correcting (FEC) code
provides excellent throughput performance over broadcast channels.
Metzner proposed an ARQ protocol based on Reed-Solomon codes for broadcast channels \cite{IBRP}.
His protocol yields much higher throughputs than those of single-to-single ARQ protocols applied 
to a broadcast channel.
Chandran and Lin presented a Selective-Repeat ARQ protocol for broadcast channels \cite{SRB}.
Sakakibara and Kasahara introduced the concept of hybrid ARQ based on GMD decoding to 
Metzner's protocol and reported improved throughput performance \cite{MHA}.
Recently, growing demand for real-time multicast, such as video streaming,
have stimulated research on FEC-based protocols based on 
Reed-Solomon codes or sparse graph codes.
For example, digital fountain codes \cite{LT} \cite{RAP} provide high throughput performance 
without introducing complex encoding/decoding operations at the server and receiver sides \cite{RCRMDD} \cite{RMDD}.

The concept of {\em index coding} proposed by Bark and Kol \cite{ISCOD} has had a significant impact on research into coding for broadcast channels. Index coding is a coding technique to achieve better bandwidth efficiency of a broadcast channel. Index coding can use side-information to improve the throughput of the protocol. The concept of index coding is described as follows. Let us assume a broadcast channel with one server and multiple receivers. The server has a packet sequence, and each receiver knows a part of the packet sequence, which is referred to as side-information. A receiver is assumed to have its own need for packets, i.e., each receiver wants to know a part of the packet sequence. The server perfectly knows the desired part of the packet sequence and side-information for each receiver. Based on such knowledge, the server can produce coded packets by combining the original packets, which are sent to the channel. Appropriately coded packets using index coding can satisfy all of the demands of receivers and reduce the number of packets to be sent. Theoretical aspects of linear index coding are discussed in Bar-Yossef et al. \cite{ICSI}, who showed that  the {\em minrank}  of the side-information graph gives the shortest code length of linear index coding. A number of theoretical studies on index coding have been conducted. For example, the relationship between index coding and network coding is discussed in \cite{ONIN}. From a practical point of view, finding appropriate combinations of packets is the most difficult part of the index coding process. Bark and Kol \cite{ISCOD} proposed a greedy type algorithm for searching large cliques in a given side-information graph. Several efficient algorithms based on a graph algorithm or SAT solver are discussed in \cite{EAIC}.

In the present paper, for reliable content distribution over 
broadcast channels with symbol (i.e., packet) erasure, we propose an ARQ protocol, referred to as {\em index ARQ protocol}, based on the concept of index coding.
The goal of the proposed scenario is to share a packet sequence  with 
all receivers participating in the protocol. 
Due to packet erasure, some of the received packets are missing at certain receivers.
The proposed protocol does not rely on conventional FEC to compensate such packet losses. 
However, unlike conventional ARQ protocols, the proposed protocol performs an encoding procedure similar to index coding to produce a retransmitted packet.
The states of the receivers are fed back to the server and such state information is used to 
make an appropriate coded packet. A coded packet is constructed 
by superimposing several packets over a finite field. This coding process resembles index coding based on the greedy clique algorithm proposed by Bark and Kol \cite{ISCOD}. Successfully received packets at each receiver 
act as side-information, and these packets can be used to improve the bandwidth efficiency of the system.
In the proposed protocol, one packet may compensate several packet losses at several receivers.
Another notable feature of the proposed protocol is that the decoding process at the receiver side has low decoding complexity because a small number of addition operations are needed in order to recover an intended packet.
This feature may be preferable for reducing power consumption in receivers.

\section{Preliminaries}
In this section, we introduce the notation and definitions used throughout the present paper.

\subsection{Broadcast channel}
\begin{figure}[t] 
\centering
\includegraphics[width=.47\textwidth]{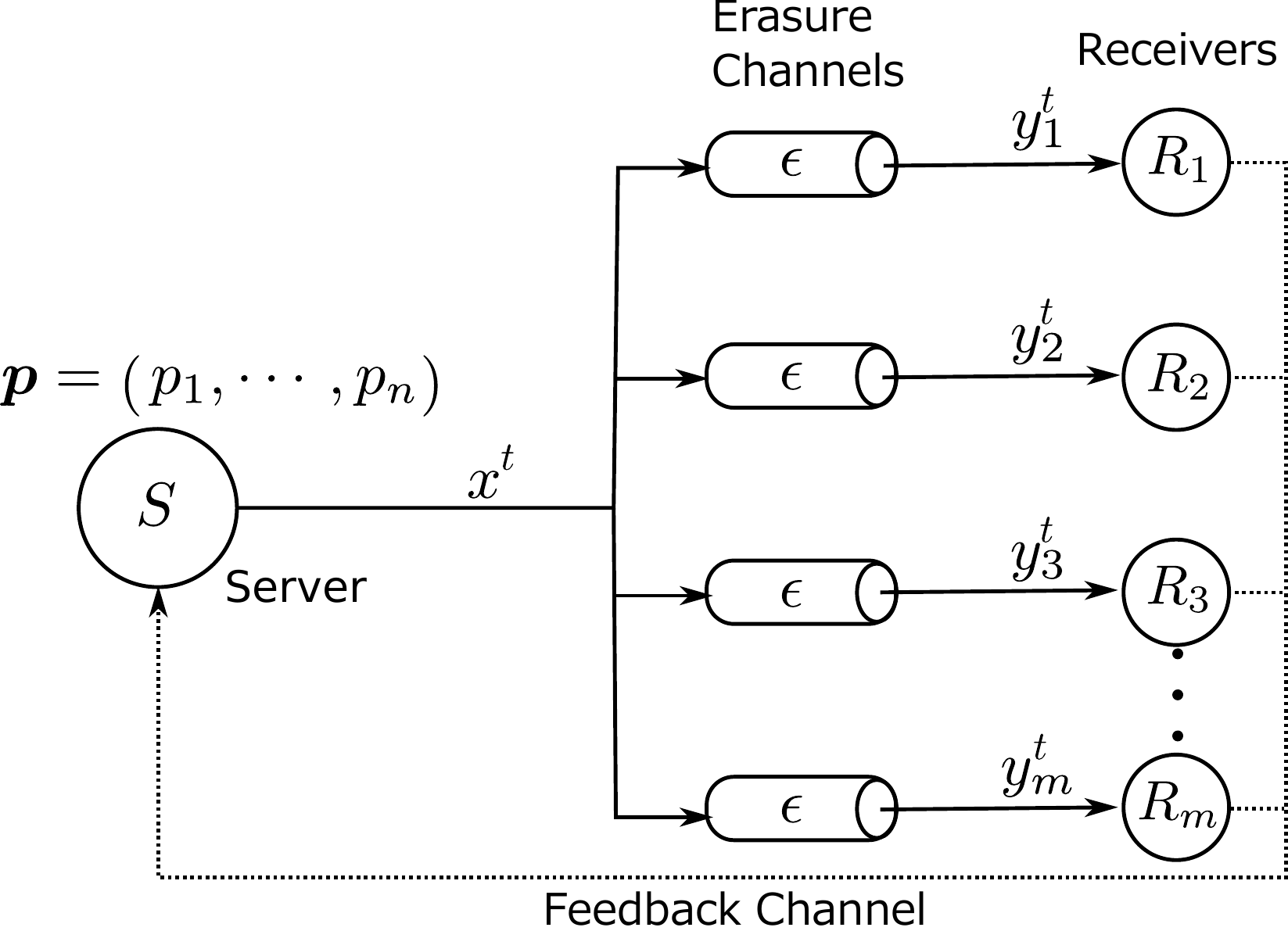} 
\caption{Broadcast channel with symbol (packet) erasure}
\label{model}
\end{figure}

Figure \ref{model} represents the broadcast channel assumed in the present paper.
A server $S$ wishes to share a packet sequence $\bm{p}=(p_1,p_2,\ldots, p_n) \in \Bbb F_q^n$ 
with $m$ receivers $R_1,\ldots, R_m$.
The symbol $\Bbb F_q$ denotes the finite field with $q$ elements, where $q$ is a prime power.
An element in $\bm{p}$, $p_i$, is said to be a {\em packet}.
Each receiver wishes to obtain whole packets in $\bm{p}$. In other words, the goal of communication over this channel 
is to distribute $\bm{p}$ to all of the receivers.

For simplicity, we assume that the server $S$ can send 
a coded packet (or an uncoded packet) $x^t \in \Bbb F_q^n$ to the channel 
at the discrete time instant $t \in \Bbb N$, where $\Bbb N$ is the set of positive integers.
The time interval between two consecutive packet transmissions is assumed to be sufficient to 
accommodate a packet. In other words, two consecutive transmitted packets never collide with each other.
The receiver $R_i (i \in [1,m])$ receives the received packet $y_i^t = x^t$ with 
probability $1-\epsilon (0 < \epsilon < 1)$;
otherwise $y_i^t = E$ with probability $\epsilon$, where the symbol $E$ represents the erasure symbol.
An erasure can be considered as the occurrence of a packet loss on the channel.
The occurrences of erasures (i.e., packet losses) are assumed to be independent (i.e., the channel is memoryless).
The notation $[a,b]$ represents the set of consecutive integers from $a$ to $b$.

In the initial state of the protocol, 
none of the receivers have knowledge of the contents of the packet sequence $\bm{p}$.
The sever continues to send a sequence of coded packets $x^1, x^2, x^3, \ldots $ 
until all of the receivers successfully obtain all of the packets in $\bm{p}$.
At time $t$, the packet indices corresponding to the packets that were successfully received by $R_i$ are denoted by $\mathcal{K}_i^t \subset [1,n]$ (known indices).
The indices of unknown packets are represented as $\mathcal{W}_i^t \subset [1,n]$ (wanted indices).
Based on these definitions, $\mathcal{K}_i^t \cup\mathcal{W}_i^t=[1,n]$ holds 
for any $i \in [1,m]$ and for any $t \in \Bbb N$.

For any time $t$, the information on $\mathcal{W}_i^t$ (or equivalently $\mathcal{K}_i^t$) is fed back to 
the sever via a noiseless feedback channel
before $x^{t+1}$ is sent to the channel. 
In other words, the server $S$ always has perfect 
knowledge of the known packets for all receivers.
Here, we assume that the size of a packet is much larger than the index information 
communicated via the noiseless feedback channel. This means that 
the capacity of the feedback channel can be much smaller than that of the forward broadcast channel.
For example, in a multicast scenario, the noiseless feedback channel can be implemented 
using reliable TCP connections.

\subsection{State matrix}

As described in the previous section, the server $S$ perfectly knows the states of the receivers.
The state matrix is a matrix representation of the knowledge of the server.
The definition is given as follows.
\begin{defi}[State matrix]
An $m \times n$ binary matrix $C=\{C_{i,j}\}$ is said to be a state matrix 
if $(i,j)$ element $C_{i,j} (i \in [1,m], j \in [1,n])$ is given by
\begin{equation}
C_{i,j} = 
\left\{
\begin{array}{ll}
1, & R_i \mbox{ knows the content of packet } p_j,  \\
0, & \mbox{otherwise}.
\end{array}
\right.
\end{equation}
\end{defi}
If the context requires the time instant (or time index) to be specified, we will use the notation $C^t$, which clarifies the dependency on the time index $t$. Otherwise, we omit the time index in order to simplify the notation.
A row of a state matrix corresponds to a receiver, and a column corresponds to a packet.
For example, assume that the server $S$ has a packet sequence $\bm{p}=(p_1,p_2,p_3)$ and 
wants to distribute the sequence to two receivers $R_1,R_2$. At a certain time, the state matrix is given by 
\begin{equation}
\label{exC}
C=\left[ 
\begin{array}{ccc}
0 & 0 & 1 \\
1 & 1 & 0 \\
\end{array} 
\right],
\end{equation}
which represents the state of the entire system. In this case,
the receiver $R_1$ knows packet $p_3$, and receiver $R_2$ knows packets $p_1$ and $p_2$.

\subsection{Clique matrix}

Let $I=( i_1,i_2,\ldots, i_\ell ) \subset [1,n]$ be an index sequence (i.e., ordered set)
satisfying $i_1 < i_2 < \cdots < i_\ell$, where $\ell (\le n)$  is a positive integer.
Assume that a state matrix 
$C = (\bm{c}_1, \bm{c}_2,\ldots, \bm{c}_n) (\bm{c}_i \in \{0,1\}^m )$ is given, 
where $\bm{c}_i$ represents the $i$th column vector of $C$.
The submatrix of $C$ indexed by $I$, which is denoted by $C_I$, is defined as
\begin{equation}
C_I =  (\bm{c}_{i_1},\bm{c}_{i_2},\dots,\bm{c}_{i_\ell}).
\end{equation}

The following definition describes clique matrices that play an important role 
in encoding and decoding processes of the proposed protocol.
\begin{defi}[Clique matrix]
If an $s \times r$ binary matrix $A$ satisfies the following two conditions: 
(1) every row of $A$ has a weight greater than or equal to $r-1$;
(2) $A$ does not contain a column with column weight $s$,
then the matrix $A$ is said to be a {\em clique matrix}.
\end{defi}
For a given state matrix $C$, if an index sequence $I$ provides a clique matrix $C_I$, 
then $I$ is called a set of  {\em clique indices} of $C$.
The term ``clique matrix'' comes from the clique based-index coding method 
presented by Bark and Kol \cite{ISCOD}. A state matrix can be seen as the 
adjacency matrix of a side-information graph. Under such an interpretation, 
a clique matrix corresponds to a clique in a given side-information graph.
Next, we present an example. Assume that the system has the state represented by (\ref{exC}).
A sub-matrix indexed by $I=(1,3)$
\begin{eqnarray}
  C_I=\left[
     \begin{array}{cc}
         0 & 1 \\
         1 & 0 \\
     \end{array}
  \right]
  \label{A}
 \end{eqnarray}
is a clique matrix. Therefore, $I=(1,3)$ is a set of clique indices in this case.

\section{Index ARQ protocol}

\subsection{Encoding process of index ARQ protocol}

We assume that the server $S$ can send a coded packet to the channel at any 
time index until all of the receivers successfully obtain the entire packet sequence.
The encoding process of the index ARQ at the server side is summarized as
{Algorithm \ref{alg1}}.

\begin{algorithm}                      
\caption{Encoding process at the server side}         
\label{alg1}                          
\begin{algorithmic}[1]                  
\STATE $t:=0$.
\STATE $C^t:= 0^{m \times n}$. ($0^{m \times n}$ represents $m \times n$ zero matrix)
\WHILE {$C^t$ contains a zero element}
\STATE 
\begin{equation} \label{indexcomp}
I^t := f^t(C^t).
\end{equation}
\STATE A coded packet is constructed as
\begin{equation} \label{encodingrule}
x^t := \sum_{j\in I^t } p_j.
\end{equation}
\STATE Send $x^t$ to the broadcast channel. 
\STATE $t := t + 1$.
\STATE $C^{t}$ is obtained based on the feedback information.
\ENDWHILE
\end{algorithmic}
\end{algorithm}

The key of the encoding process is (\ref{indexcomp}) and (\ref{encodingrule}) in {Algorithm \ref{alg1}}. 
We here focus on these steps.
The function $f^t: \Bbb F_q^{m \times n} \rightarrow 2^{[1,n]}$ finds a set of clique indices 
from a given state matrix. In other words,  $C_{I}$ is a clique matrix in $C$, where $I = f^t(C)$.
The function $f^t$ is referred to as an {\em index generator}.
The details of an implementation of an index generator are discussed in Subsection \ref{indexgen}.
The encoding process at the server side continues until the state matrix $C$ has no 
elements with the value one.

According to (\ref{encodingrule}), 
a coded packet $x^t \in \Bbb F_q$ is encoded by adding packets having indices in $I$.
Several packets are superimposed over $\Bbb F_q$ to produce a coded packet.
This coded packet can thus satisfy the demands from several receivers simultaneously, i.e., a coded 
packet can compensate multiple unknown packets in several receivers. This provides the advantage of the proposed protocol in terms of throughput and bandwidth efficiency.

\subsection{Decoding process of index ARQ protocol}

\begin{figure}[t] 
\centering
\includegraphics[width=.47\textwidth]{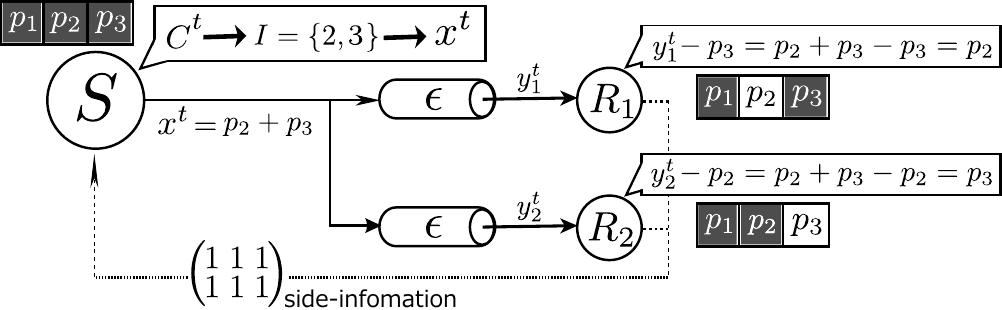} 
\caption{Encoding and decoding processes of the index ARQ protocol.
The server broadcasts a coded packet $x^t$. The receiver $R_1$ 
obtains $y_1^t$ and attempts to retrieve the packet $p_2$ using the side-information.
In a similar manner, the receiver $R_2$ can recover intended  packets.}
\label{comflow}
\end{figure}

The receiver $R_i$ receives  symbol $y_i^t$ from the channel.
We assume that each receiver knows $I^t$ via the header information attached to a 
transmitted coded packet.
The decoding process at time index $t$ in the receiver $R_i$ is summarized 
in Algorithm \ref{alg2}.                          
\begin{algorithm}                      
\caption{Decoding process at the receiver $R_i$}         
\label{alg2}                          
\begin{algorithmic}[1]                  
\IF{$y_i^t = E$}
\STATE Quit the decoding process. 
\ENDIF
\IF {$I^t \cap {\cal W}_i^t = \emptyset$}
\STATE Quit the decoding process. 
\ELSE
\STATE Let $k$ be a unique index in $I^t \cap {\cal W}_i^t$.
\ENDIF
\begin{equation} \label{reconst}
    p_k := y^t_i - \sum_{j\in I^t \backslash \{k\}} p_j.
\end{equation}
\STATE ${\cal W}_i^{t+1} := {\cal W}_i^t \backslash \{k\}$.
\STATE Send ${\cal W}_i^{t+1}$ to the server via the feedback channel.
\end{algorithmic}
\end{algorithm}
The packet $p_k$ is an unknown packet before decoding because $k \in {\cal W}_i^t$.
Note that $R_i$ knows the packet $p_j$ for any $j\in I^t \backslash \{k\}$
due to the definition of the clique matrix and clique indices.
The reconstruction rule (\ref{reconst}) is 
an immediate consequence of the encoding rule (\ref{encodingrule}).
As an example, encoding and decoding processes are depicted in Fig.~\ref{comflow}.

\subsection{Index generator}
\label{indexgen}

As described in the previous subsections, a coded packet consists of several original packets. In order to improve the bandwidth efficiency, we need to increase the number of superimposed packets in an encoding process. This means that an index generator that is able to find a larger clique matrix from a state matrix is a preferable choice. In this subsection, we will present such an index generator.

In the present paper, we use the following strategy to design an index generator.
The {\em first phase} of the protocol is defined as a sequence of time indices $(1,2,\ldots, n)$.
The index generator outputs $\{t \}$ if $t$ is in the first phase, i.e., $t \in [1,n]$.
In other words, the original packets $p_1,p_2,\ldots, p_n$ are sent directly to the channel in the first phase. The {\em second phase} of the protocol ($t > n$) can be considered to be a retransmission phase. In the second phase, unreceived packets due to packet losses are gradually compensated as the decoding process proceeds. The index generator used in the second phase is based on a greedy algorithm to find a large clique matrix. The problem of finding the largest (in terms of the number of columns) clique matrix is closely related to the problem of finding the largest clique in a given undirected graph (maximal clique problem). As in the case of the maximal clique problem, we cannot expect the existence of efficient algorithms for this problem. In the present paper, a heuristic greedy algorithm for finding a clique matrix is used in the second phase. In summary, our index generator has the following form:
\begin{equation}
f^t(C) = 
\left\{
\begin{array}{cc}
\{t \}, & t \in [1,n] \\
F(C), & \mbox{otherwise},
\end{array}
\right.
\end{equation}
where the function $F$ represents a greedy process to compute clique indices.

The greedy search algorithm presented below is a randomized greedy algorithm 
for finding a set of clique indices. This algorithm is used to implement the function $F(C)$.

\begin{algorithm}                      
\caption{Greedy search algorithm}         
\label{alg3}                          
\begin{algorithmic}[1]                  
\STATE $S := \{\bm{c}_1, \bm{c}_2,\ldots, \bm{c}_n \}$ ($C = (\bm{c}_1, \bm{c}_2,\ldots, \bm{c}_n)$).
\STATE $A := ()$. 
\STATE $I := \emptyset$.
\WHILE{$S\neq\emptyset$} 
\STATE Select a vector $\bm{c}_k$ from $S$ uniformly at random.
\IF {$\bm{c}_k$ is not the all-ones vector}
\STATE  $A' := (A, \bm{c}_k)$.
\IF {$A'$ is a clique matrix}   
\STATE $A := A'$.
\STATE $I := I \cup \{k\}$.
\ENDIF
\ENDIF
\STATE $S := S \backslash \{\bm{c}_k\}$.
\ENDWHILE
\STATE Output $I$.
\end{algorithmic}
\end{algorithm}
The concept of the algorithm is simple. 
In line 6 of Algorithm \ref{alg3}, a column of the state matrix $\bm{c}_k$ is randomly chosen 
and is appended to the current candidate of clique matrix $A$ if $(A, \bm{c}_k)$ does not violate 
the condition of the clique matrix. This process continues until all of the columns in $C$ are tested.
The overall time complexity of this greedy search is $O(n^2)$ when $n = m$.
The randomness on the column selection is incorporated  because it provides robust 
system performance with regard to the delay of feedback information.

Next, we consider a simple example. Assume that we have the state matrix $C$, as in (\ref{exC}), and
that the order of the random column choice is second column $\rightarrow$ first column $\rightarrow$ third column. In such a case, the second column is first accepted and the first column is rejected because it forms a non-clique matrix. Finally, the third column is accepted, and the algorithm outputs $I=(2,3)$. 

Although this algorithm depends on a simple greedy strategy, the greedy algorithm has been empirically observed to produce clique indices near the optimal (i.e., largest) size.

\section{Computer experiments}

\subsection{Details of experiments}

In this subsection, we describe the details of computer experiments for evaluating the throughput performance of the proposed protocol.
We assume the broadcast channel shown in Fig. \ref{model}.
As benchmarks, we evaluate the performance not only of the proposed protocol 
but also of the Selective-Repeat (SR) protocol and the Metzner protocol.

The SR protocol may be the simplest ARQ protocol for compensating the packet loss under this 
channel model. The details of the SR protocol are as follows. As in the proposed protocol, 
the server transmits the original packet $p_t$ when $t = 1,2,\ldots, n$ (phase 1). At every time interval, each receiver reports its demands (i.e., state of the unreceived packets) to the sever via the reliable feedback channel. In phase 2 of this protocol, the packet with the smallest index among all of the requested packet indices is sent to the channel at time index $t > n$. When the server receives ACK from all of the receivers, the server terminates the transmission process. The advantage of the SR protocol is its simplicity. No special operations are required for encoding and decoding. However, the SR protocol cannot provide a coding advantage to improve the throughput.

The Metzner protocol is an FEC-based ARQ protocol for broadcast channels that is based on the erasure correcting capability of Reed-Solomon codes. In the Metzner protocol, Reed-Solomon coded packets are sent to the channel, and a receiver that obtains $n$ packets from the channel can execute an erasure correcting process (i.e., solving simultaneous linear equations over $\Bbb F_q$) to recover the packet sequence that the server possesses. The primary benefit of this protocol is the near-optimal bandwidth efficiency it provides.
A drawback of the protocol is that every receiver requires to solve an erasure correcting problem 
based on a Reed-Solomon code that requires a certain computational power at the receiver side.

In the present paper, we adopt {\em throughput} as the main performance measure.
The throughput of a protocol is directly related to the bandwidth efficiency of the protocol.
Let $N$ be the total number of transmitted packets from the server until the protocol terminates 
(i.e., all of the receivers obtain the entire packet sequence). Note that this number $N$ is 
a random variable depending on the randomness of the erasure channels.
The throughput $\tau$ is defined as
$
\tau = {\sf E} \left[{n}/{N} \right]
$
which represents average amount of information per transmitted packet.
In the following subsections, the throughput $\tau$ is estimated through computer simulations of
these protocols.

The capacity of a single erasure channel with erasure probability $\epsilon$ is given by $1-\epsilon$.
The coding theorem proved by Shannon \cite{EIT} guarantees the existence of sufficiently long FEC-codes with coding rates below $1-\epsilon$ that achieve arbitrarily small error probabilities. Suppose that a server uses such an FEC-code with a coding rate close to $1-\epsilon$. Although such a system suffers from large latency due to long code length, a throughput close to $1-\epsilon$ can be achieved. In the following discussion, we use such a system as a benchmark for the throughput performance. The upper bound of the throughput $\tau = 1 - \epsilon$ is referred to hereinafter as the {\em ideal FEC} bound.

\subsection{Results of computer experiments}

\begin{figure}[t] 
\centering
\includegraphics[width=.48\textwidth]{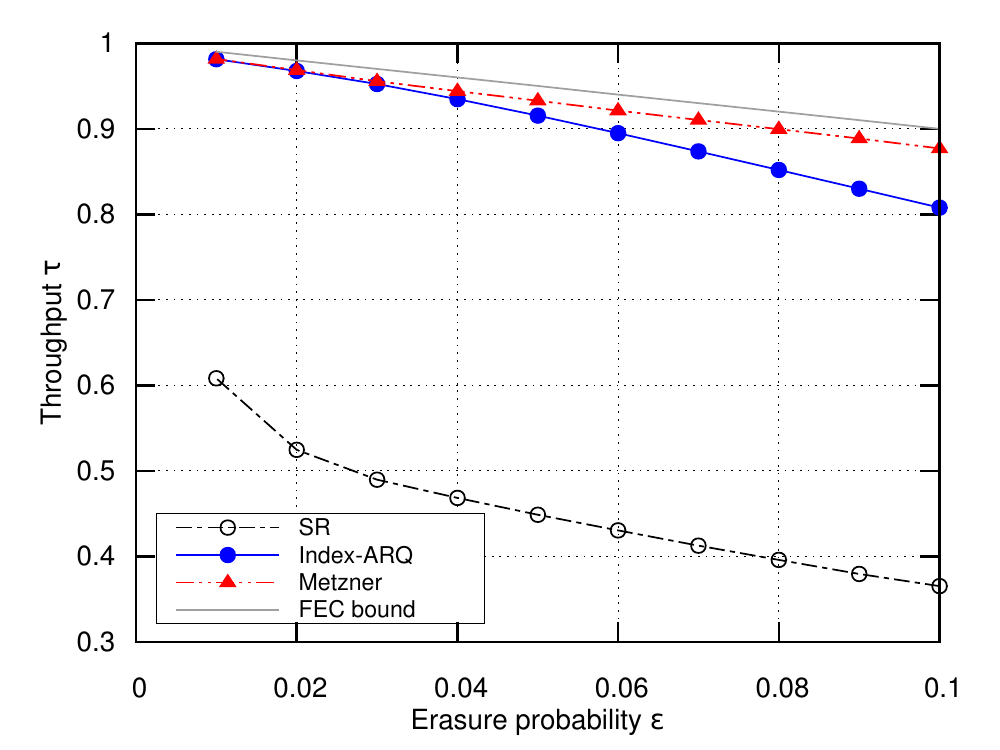} 
\caption{Relationship between throughput and erasure probability ($\epsilon: 0\leq\epsilon\leq 0.1$, number of receivers $m=100$, number of packets  $n=1000$).}
\label{erasureChange}
\end{figure}

Figure \ref{erasureChange} shows the relationship between throughput 
and erasure probability $\epsilon$ ($0 \leq \epsilon \leq 0.1$).
In these experiments, 100 trials were conducted for each point on the curves.
The number of receivers and packets are assumed to be $m=100$ and $n=1000$, respectively.
The results for three protocols, including the proposed protocol (labeled by index-ARQ), the selective repeat (SR) protocol, and the Metzner protocol, are included in Fig. \ref{erasureChange}.

\begin{figure}[t] 
\centering
\includegraphics[width=.48\textwidth]{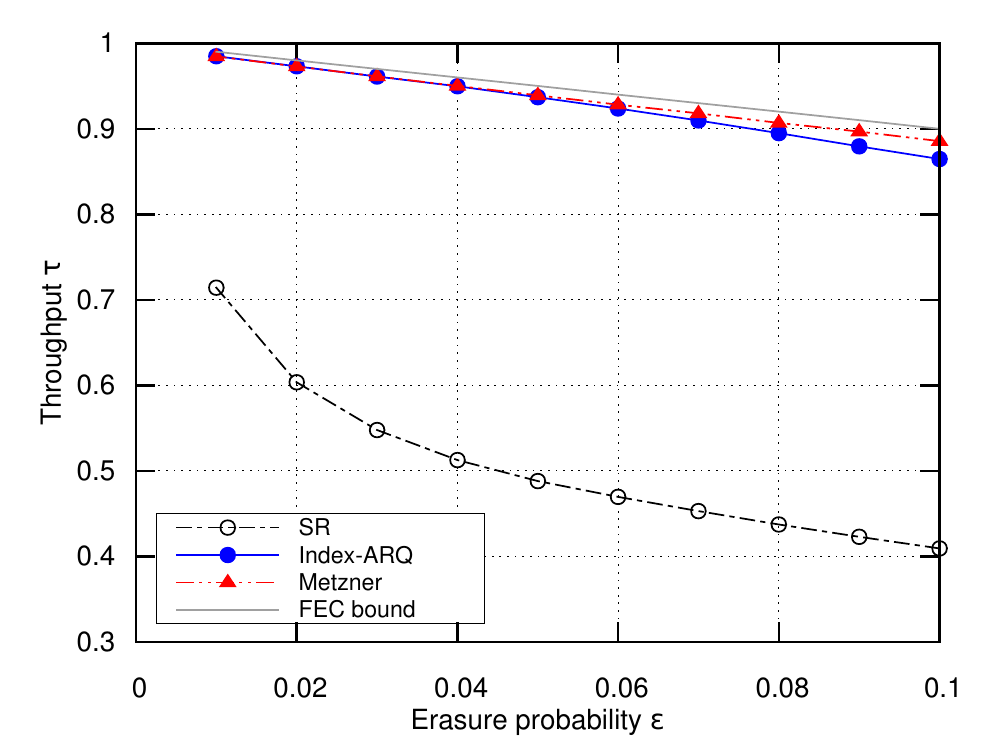} 
\caption{Relationship between throughput and erasure probability ($\epsilon: 0\leq\epsilon\leq 0.1$, number of receivers $m=50$, number of packets  $n=2000$).}
\label{erasureChange2}
\end{figure}

The Metzner protocol is confirmed to achieve the best throughput performance among these three protocols at all erasure probabilities. The throughput of the Metzner protocol are quite close to the ideal FEC bound $\tau = 1-\epsilon$.
This fact indicates that the Metzner protocol provides excellent bandwidth efficiency that is close to optimal. On the other hand, the SR protocol offers poor throughput performance compared with the Metzner protocol. 
For example, the Metzner protocol provides $\tau = 0.877$, whereas the SR protocol yields $\tau = 0.365$ at $\epsilon = 0.1$.
This result implies that the SR protocol cannot achieve a bandwidth efficiency close to the optimal performance, although it is the simplest to implement.
The proposed protocol, index-ARQ, provides slightly smaller throughput compared with the Metzner protocol but 
the difference is fairly small, especially when the erasure probability is small, such as $\epsilon < 0.05$.
Even for a relatively large erasure probability $\epsilon = 0.1$, the proposed protocol achieves $92\%$ of the throughput performance of the Metzner protocol.

Figure \ref{erasureChange2} show the case in which $m=50$ and $n=2000$.
In this case, we can also observe the same tendency seen in Fig.~\ref{erasureChange}.

\begin{figure}[t] 
\centering
\includegraphics[width=.48\textwidth]{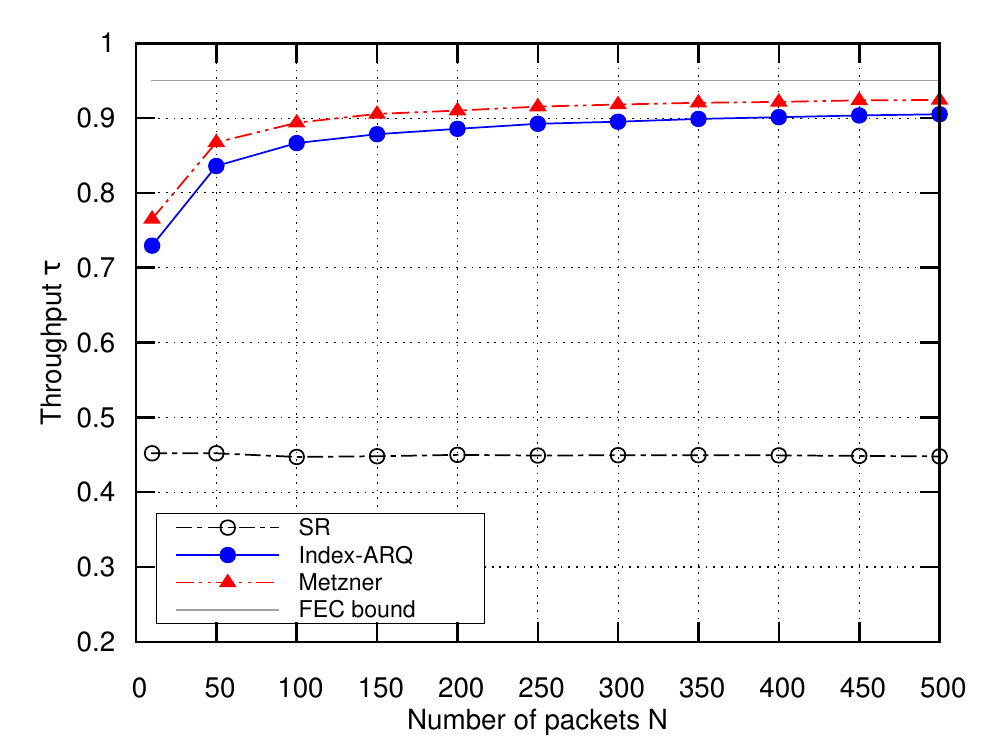} 
\caption{Relationship between the number of packets ($n\in[10,500]$) and throughput ($\epsilon=0.05$, $m=100$)}
\label{receiverConst}
\end{figure}

In order to observe the relationship between the number of packets and the throughput, we conducted several experiments. Figure \ref{receiverConst} shows such a relationship under the condition in which the erasure probability is $\epsilon=0.05$ and the number of receivers is $m=100$. The horizontal axis indicates the number of packets, and the vertical axis indicates the throughput.
In the case of the SR protocol, the throughputs are approximately constant, regardless of the number of packets. On the other hand, in the case of the index ARQ protocol and the Metzner protocol, 
the throughputs increase slightly as the number of packets increases.
Moreover, the difference in throughput of these two protocols and the ideal FEC bound
becomes negligible as the number of packets increases.
The experimental results suggest a system design principle for the index ARQ protocol such that 
the number of packets should be 
larger than the number of receivers in order to achieve higher throughput.

\section{Concluding summary}

In the preset paper, we proposed 
an ARQ protocol, referred to as index ARQ, for broadcast channels.
In the proposed protocol, the server incorporates a coded packet by adding several packets over $\Bbb F_q$ based on the knowledge 
of the demands of all of the receivers. In order to find an appropriate set of indices for the packets to be added, a randomized greedy algorithm is devised. The bandwidth efficiency of the proposed protocol derives from 
the fact that a coded packet can compensate multiple packet losses in several receivers.
A practical advantage of the proposed protocol is the simplicity of its decoding process at the receiver side. A decoding process only constitutes several additions over $\Bbb F_q$ and results in a small computational load at the receiver side. If $q = 2^m$, then only exclusive OR operations are required to recover packet losses.

Based on the results of computer experiments, we confirmed that the proposed protocol achieves much higher throughputs than the SR protocol. The proposed protocol requires a certain computational load to encode at the server side. This computational load at the server side can be considered as a cost to be paid in order to achieve better bandwidth efficiency than the SR protocol.
The throughput performance of the proposed protocol is close to that of the Metzner protocol and the ideal FEC bound when the erasure probability is in the range of $0 < \epsilon < 0.1$, which implies that the proposed protocol provides approximately optimal throughput performance in such a regime. If a receiver is a mobile terminal with lower computational power or prefers a low power consumption, the proposed protocol would be a preferable choice in order to achieve both high bandwidth efficiency and lower computation load at the receiver side.

\section*{Acknowledgement}
The present study was supported by a Grant-in-Aid for Scientific Research (B) (Number 25289114) from JSPS.


\begin{thebibliography}{15}
\bibitem{IBRP} J. J. Metzner, ``An improved broadcast retransmission protocol,'' IEEE Trans. Commun., vol.32, pp. 679-683, 1984.

\bibitem{SRB} S. R. Chandran and S. Lin, ``Selective-repeat-ARQ schemes for broadcast links,'' IEEE Trans. Commun, vol. 40, no. 1, pp. 12-19, Jan. 1992.

\bibitem{MHA} K. Sakakibara and M .Kasahara, ``A multicast hybrid ARQ scheme using MDS codes and GMD decoding,'' IEEE Trans. Commun., vol.43, pp. 2933-2940. 1995

\bibitem{RCRMDD} M. Luby, T. Gasiba, T. Stockhammer, M. Watson, and W. Xu, ``Raptor codes for reliable download delivery in wireless broadcast systems,'' IEEE Consumer Comm. and Networking Conf, vol. 1, pp. 192-197. Jan. 2006

\bibitem{RMDD} M. Luby, T. Gasiba, T. Stockhammer, and M. Watson, ``Reliable multimedia download delivery in cellular broadcast networks,'' IEEE Tran. Broadcasting, vol. 53, no. 1, pp. 235-246, Mar. 2007.

\bibitem{PLRMT} J. Nonnenmacher, E. W. Biersack, and D. Towsley, ``Parity-based loss recovery for reliable multicast transmission,'' Proceerings of ACM SIGCOMM. Sept. 1997.

\bibitem{LT} M. Luby,  ``LT codes,'' Proc. Annu. Symp. Found. Comput. Sci. Vancouver, Canada, 2002, pp. 271-282. 

\bibitem{RAP} A. Shokrollahi, ``Raptor codes,'' IEEE Trans. Inf. Theory, vol. 52, no. 6, pp. 2551-2567, June. 2006. 

\bibitem{ISCOD} Y. Bark and T. Kol, ``Informed-source coding-on-demand (ISCOD) over broadcast channels,'' In Proceeings of INFOCOM'98, 1998.

\bibitem{ICSI} Z. Bar-Yossef, Y. Birk, T. S. Jayram, and T. Kol, ``Index coding with side information,'' IEEE Trans. Inf. Theory, vol. 52, no. 6, pp.1479-1494, Mar. 2011.

\bibitem{ONIN} S. E. Rouayheb, A. Sprintson, and C. Georghiades, ``On the relation between the index coding and the network coding problems,'' in Proc. IEEE Symp. Inf. Theory, Toronto, Canada, 2008, pp. 1823--1827.

\bibitem{EAIC} MAR. Chaudhry, and A. Sprintson, ``Efficient algorithms for index coding,'' INFOCOM Workshops 2008, pp. 1-4.

\bibitem{FLC} M. M. Ali, and U. Niesen, ``Fundamental limits of caching,'' IEEE Trans. Inf. Theory, vol. 60, no. 5, pp. 2856--2867, May 2014.

\bibitem{BWSI} N. Alon, E. Lubtzky, U. Stav, A. Weinstein, and A. Hassidim, ``Broadcasting with side information,'' in 49th Ann. IEEE Symp. Found. Comput. Sci., Philadelphia, PA, Oct. 2008, pp. 823--832.

\bibitem{EIT} T. M. Cover and J. A. Thomas. \textit{Elements of information theory.} John Wiley \& Sons, 2012.

\end{thebibliography}
\end{document}